# FMEA Based Risk Assessment of Component Failure Modes in Industrial Radiography

Alok Pandey[#1], Meghraj Singh[#2], A. U. Sonawane[#3], Prashant S. Rawat*[4]

[#]*Atomic Energy Regulatory Board, Niyamak Bhavan, Anushakti Nagar, Mumbai 400094, India*
[*]*Dept. of Nuclear Science & Technology, University of Petroleum and Energy Studies, Bidholi, Dehradun 248007, India*

***Abstract -*** Industrial radiography has its inimitable role in non-destructive examinations. Industrial radiography devices, consisting of significantly high activity of the radioisotopes, are operated manually by remotely held control unit. Malfunctioning of these devices may cause potential exposure to the operator and nearby public, and thus should be practiced under a systematic risk control. To ensure the radiation safety, proactive risk assessment should be implemented. Risk assessment in industrial radiography using the Failure Modes & Effect Analysis (FMEA) for the design and operation of industrial radiography exposure devices has been carried out in this study. Total 56component failure modes were identified and Risk Priority Numbers (RPNs) were assigned by the FMEA expert team, based on the field experience and reported failure data of various components. Results shows all the identified failure modes have RPN in the range of 04 to 216 and most of the higher RPN are due to low detectability and high severity levels. Assessment reveals that increasing failure detectability is a practical and feasible approach to reduce the risk in most of the failure modes of industrial radiography devices. Actions for reducing RPN for each failure mode have been suggested. Feasibility of FMEA for risk assessment in industrial radiography has been established by this study.

## I.  INTRODUCTION

Industrial radiography is an important non-destructive evaluation technique in which radioisotopes (generally referred as source) are used for imaging of weld joints and castings, for detection of any flaw. Devices used for industrial radiography operations provides shielding to the source during storage and therefore reduces the ionizing radiations to the permitted levels. Radiography device consists of various mechanical components, to provide the safety systems and source transition mechanism. Functionality of these components are crucial, as their failure may cause potential radiation exposure to operating team and nearby public. The dose rates that prevail close to a source or a device may be high enough to cause overexposure of extremities within a matter of seconds, and can result in the loss of a limb. Throughout the history of industrial radiography, accidents with some sources have occurred that have resulted in injuries [1].

As failure of radiography device or its components for its intended function may cause untoward accidents involving high exposure to ionizing radiation, it is essential to ensure that these devices are equipped with the necessary safety features for operation. Regulatory agencies of respective country carryout the safety assessment of the new and existing models of radiography devices, based on the design safety requirements stipulated in the international [2]/national standards and operational feedback from the operators. Design and safety features of the radiography devices has changed significantly with time, from manually operated shutter-type devices in 1990s to current version of remotely operated devices. To enhance the operational safety, several advance features have been added in the design like rotation of selector ring, colour indicator for source location etc.

Malfunctioning of the radiography devices has been identified as initiating event for accidents which has resulted in the deterministic health effects to the operator and public [3].With advancement in the design of the exposure devices, the incidents and the effective dose to the operating personnel has reduced. Risk assessment for these devices may provide inputs for further advancements in the design to enhance the radiation safety. Application of risk assessment methodologies like probabilistic safety assessment and Failure Modes & Effects Analysis (FMEA) for risk assessment in non-reactor radiation/nuclear facilities has been emphasized and encouraged by the International Commission on Radiological Protection [4] and International Atomic Energy Agency [5,6].

FMEA is a well-established and systematic approach to identify and understand contributing factors of potential failures on a process, design or practice. The main objective of FMEA is to identify potential failure modes, evaluate the causes and effects of different component failure modes, and determine what could eliminate or reduce the chance of failure [7].This risk assessment tool may be utilised to study and prioritize the consequences and the frequency of occurrence of failures associated with a system. FMEA is used for risk assessment during conceptual design and during process control after actual development of the system, as a component of continuous development [8].

FMEA is helpful to identify failure scenarios, i.e. potential accident initiators. At the level of individual systems, FMEA may be useful in identifying failure contributions to be modelled in fault trees[5].This risk assessment methodology has been adopted for risk assessment in various industries like nuclear, automotive, aerospace, healthcare/medical industries etc. [9-11].However, limited work has been published for application of FMEA in the industrial and medical applications of radioactive sources. FMEA study has been carried out in Cuba with the objective of the safety evaluation of the performance of cobalt teletherapy by the oncological unit of Pinar del Rio (UOPR) in Pinar del Rio, Cuba, in a systematic, exhaustive and structured way[5]. Giardina et al. carried out risk assessment of component failure modes in brachytherapy using FMEA [12].Scorsetti M. et al. conducted FMEA study in radiotherapy process, practices in an Italian hospital and suggested organizational and procedural corrective measures [13].Marefat et al. carried out FMEA to compare the reliability of industrial radiography and phased-array ultrasonic testing for detection and identification of weld defect[14]. No FMEA study could be found published for risk assessment associated with the design and operation of industrial radiography exposure devices.

This study aimed to the risk assessment in industrial radiography, using FMEA technique to determine the risk priority numbers for the component failures of the radiography devices. Present study is helpful to verify the feasibility of FMEA methodology for risk assessment in industrial radiography, which is still unexplored. Results in terms of RPN identify the components which are riskier from the operational hazard point of view. Actions for improvements in the design and operating conditions have been suggested for the risk management.

## II. DESIGN AND OPERATION OF THE INDUSTRIAL RADIOGRAPHY DEVICE

Industrial radiography device servesthree purposes (a) operation for radiography exposures (b) providing shielding during non-operation time and, (c) transport container for the transport of source contained in it. Basic design and components of all the available models of these devices are same, with minor variation in the shape of shielding material housing and components of the safety systems.

The radiography device comprises of four detachable sub-units namely, source housing, remote control, guide tube(s) and source assembly. During non-operation hours, source assembly rests inside exposure container, which provides shielding to the source. For radiography operations, remote control and guide tube(s) are connected to the exposure container from the rear and front end respectively. Remote control consists of metallic control cable which provides meansforconnecting the source assembly inside exposure container. Once control cable is connected with the source assembly, handle provided in the remote control is rotated, which in turn pushes the source assembly forward in the projection sheath attached to the front end of the exposure container, until source assembly reaches at the end point containing metallic snout (exposure head). When the exposure time is completed, source is retracted back inside the source housing by rotating the handle of control unit in the reverse direction. Schematic diagram of industrial radiography exposure device is shown in figure 1.

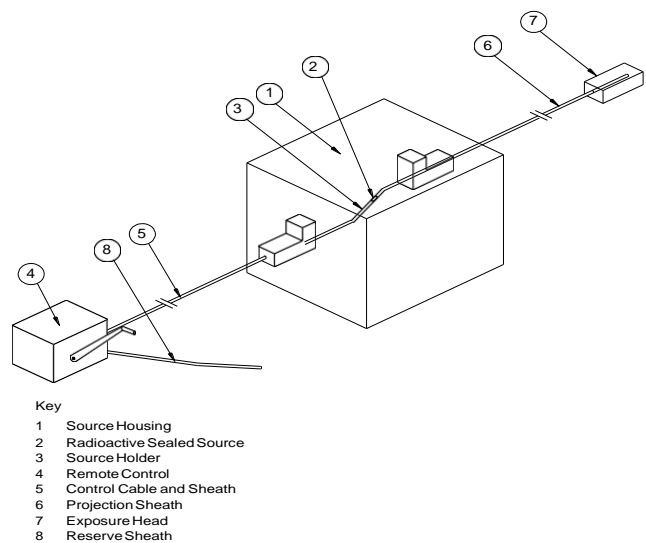

Key
1 Source Housing
2 Radioactive Sealed Source
3 Source Holder
4 Remote Control
5 Control Cable and Sheath
6 Projection Sheath
7 Exposure Head
8 Reserve Sheath

Figure 1: Industrial radiography exposure device
(Image source: ISO-3999; 2004)

## III. FMEA IN INDUSTRIAL RADIOGRAPHY

Several accidents associated with industrial radiography have been reported worldwide due to human error and equipment malfunctioning [1].Even though operation of these devices are simple, smooth source transition requires functioning of various components of the exposure device, and consequences of device malfunctioning may be very severe. Safety interlocks and indicators provided in the device have important role to prevent any incident/accident.

Risk assessment has been carried out using FMEA methodology for component failure modes of industrial radiography devices. One of the prerequisites of FMEA assessment is the constitution of committed team with its members having strong knowledge and experience of the system under study. FMEA team, comprising of ten members having experience of 10- 35 years in the respective professions, was constituted.

To provide appropriate weightage, members from all the stakeholders of radiography devices, i.e. operators, radiological safety officers, suppliers of devices/spare parts, maintenance & servicing personnel and radiation safety regulator, were included as member of the FMEA team. Before starting actual assessment, formal training about the FMEA methodology was provided to all the team members.

Study was carried out at one of the servicing and maintenance site of radiography devices. Whenever required, device with dummy source assembly was operated to simulate the actual operational conditions. Basic design and safety components of all the industrial radiography devices as same, therefore study was carried out considering a generic model of the radiography device along with the consideration of special safety provisions/interlocks provided in all commercially available models of industrial radiography devices. For the study purpose, radiography device was divided into its four sub-units namely (i) source housing, (ii) guide tube (iii) remote control unit, and (iv) source assembly. Each sub-unit was further divided upto its basic component level. Assessment outputs were compiled in a table comprising of the failure modes and effects of these failure modes on the device and the operator/public.

FMEA requires three numerical attributes for each component failure modes, (i) Occurrence of failure (O), which expresses the probability that the component failure will occur, (ii) Severity of failure (S), which expresses the severity of event resulting due to the component failure and, (iii) detection (D) which represents the probability that the incipient failure will be detected before it occurs. Risk Priority Number (RPN) is obtained as product of these indexes.

RPN= O X S X D

Numerical values of these parameters i.e. O, S & D varies from 1 to 10, which is estimated by the experts of FMEA team. Final value of RPN varies from 1 to 1000, with higher values as more critical and should be given higher priorities for correction. O,S and D ranking were assigned using standard criteria published in literatures, as given in table I-III respectively. To determine the occurrence (O) values, the failure data collected by regulatory agency during pre-source loading inspections, failure data from the records of servicing and maintenance agency, and the field experience of FMEA team members, were utilized. Severity (S) ranking were assigned on the basis of the failure effect on the person, which is the main concern of risk assessment. Severity of failure on the person has been considered as the severity of exposure to ionizing radiation. Term 'Injury' in table II, corresponds to the exposure to ionizing radiation from radioactive source.

**TABLE I. FMEA ranking for probability of Occurrence (O) for component failure [12, 15-18]**

| Probability of Occurrence | Ranking | Possible failure rate (No. of exposures) |
|---|---|---|
| Remote | 1 | < 1:20,000 |
| Low | 2 | 1:20,000 |
|  | 3 | 1:10,000 |
| Moderate | 4 | 1:2000 |
|  | 5 | 1:1000 |
|  | 6 | 1:200 |
| High | 7 | 1:100 |
|  | 8 | 1:20 |
| Very High | 9 | 1:10 |
|  | 10 | 1:2 |

**TABLE II. FMEA ranking for Severity (S) of component failure [10, 12, 19-22]**

| Effect | Rank | Severity of effect |
|---|---|---|
| No effect | 1 | No reason to expect failure. Slight annoyance- no injury to worker or public. |
| Very Minor | 2 | Very minor effect on device performance. Slight danger- no injury to worker or public. |
| Minor | 3 | Minor effect on device performance. No injury to worker or people. |
| Very Low | 4 | Very low effect on device performance. Minor or no injury to worker. |
| Low | 5 | Moderate effect on device performance. The device requires repair. Very moderate danger-minor injury to worker. |
| Moderate | 6 | Device performance is degraded. Some safety functions may not operate. The device requires repair. Moderate danger- minor to |
| | | moderate injury to worker. |
| High | 7 | Device performance is severely affected but operational with reduced level of safety performance. Dangerous-moderate to major injury to worker OR minor injury to public. |
| Very High | 8 | Primary safety function(s) of device is lost. Failure can involve hazardous outcomes. Dangerous-may result in major injury to worker OR moderate injury to public. |
| Hazardous with warning | 9 | Failure involves hazardous outcomes. Very dangerous-may result in major injury or death of worker ormajor injury to public. |
| Hazardous without warning | 10 | Failure is hazardous and occurs without warning. It suspends operation of the system. Extremely dangerous- may cause death of worker or public. |

There is no clear demarcation on the acceptable values of RPN in the literatures. Lipol et al. considers RPN as acceptable if less than 200, undesirable if between 200 and 500 and unacceptable if more than 500 [24]. Serafini et al. assumes RPN acceptable if less than 100, corrective action necessary if RPN between 100 and 150 and drastic and timely actions are necessary if RPN more than 150 [25]. For the present study, following conservative acceptance criteria for the resulted RPN was set by FMEA team, based on the experience and literature review.

(1) Acceptable if RPN ≤ 100
(2) Corrective actions recommended for 500 ≥ RPN > 100
(3) Urgent corrective actions are recommended if RPN >500

TABLEIII. FMEA ranking for Detection (D) of component failure [10, 12, 18, 22, 23]

| Detectability | Rank | Probability of detection (%) | Likelihood of detection of failure or error |
|---|---|---|---|
| Almost Certain | 1 | 86-100 | Design/operation control will almost certainly detect a potential failure mode. |
| Very high | 2 | 76-85 | |
| High | 3 | 66-75 | High chance that the design/operation control will almost certainly detect a potential failure mode. |
| Moderately high | 4 | 56-65 | |
| Moderate | 5 | 46-55 | Moderate chance that the Design/operation control will detect a potential failure mode (e.g. the defect will remain undetected until the device performance is affected). |
| Low | 6 | 36-45 | |
| Very low | 7 | 26-35 | Remote chance that the design/operation control will detect a potential failure mode (e.g. the defect will remain undetected until device inspection is carried out). |
| Remote | 8 | 16-25 | |
| Very remote | 9 | 6-15 | Defect most likely remains undetected (e.g. the design/ operation control cannot detect potential cause or the operation will be continued to be performed in the presence of the defect). |
| Absolute uncertain (impossible to detect) | 10 | 0-5 | Device/component failures are not detect (e.g. there is no design/operation verification or the operation will certainly be continued to perform in the presence of the defect) |

## IV. RESULTS

Risk assessment for the industrial radiography exposure device was carried out by dividing the whole device into its four sub-units. Each sub-unit was further divided up to component level and failure mode of each component was discussed in detail. RPNs were calculated for all the identified failure modes of each sub-unit of industrial radiography exposure device.

Total 56 failure modes were identified & assessed, and corresponding rankings were assigned. 25 failure modes which are considered important and severe, are given in table IV. Resulted RPN for failure modes varies from 04 to 216. The highest RPN obtained from the study which is 216, is much lower than the maximum possible value of RPN i.e. 1000, which reflect that very severe failure modes in the existing design of industrial radiography devices are unusual.

Occurrence ranking of most of the failure modes are on lower side only. However, the lower detectability and higher severity contributes to increase in RPN values. Reducing severity of the failure modes (which is the exposure to the ionizing radiation), is mainly dependent on the human actions, therefore to reduce the RPN of severe failures, easiest and practical way would be to increase the detection probability. FMEA team recommended the actions to reduce RPN for each of the failure modes, which are outlined in table IV. These recommended action(s) focuses mostly to increase the detection probability. Severity ranking of failure modes are provided in the last column of the table IV. In case of same RPN values, failure mode having higher severity ranking has been assigned higher RPN ranking.

**Table IV. FMEA of component failure modes in industrial radiography**

| ID | Component | Potential Failure Mode | Potential failure effect on radiography device | Failure effect on person (Occupational worker/Public) | Potential cause(s) of failure | Detection method (if any) | O | S | D | RPN (O*S*D) | Actions Recommended | Ranking |
|---|---|---|---|---|---|---|---|---|---|---|---|---|
| **Remote Control/ Driving Assembly** | | | | | | | | | | | | |
| RC1 | Control cable | Wire damaged/ broken | Source cannot be projected/retrieved | Potential exposure to occupational worker | Wear and tear | Inspection (partial wire length only) | 3 | 8 | 8 | 192 | Periodic QA testing | 2 |
| RC2 | | Male coupler dimensions worn out | Source detachment from control cable/cannot be retrieve back in the device | Potential exposure to occupational worker/public | Wear and tear | Inspection | 2 | 9 | 3 | 54 | Periodic QA testing | 10 |
| RC3 | | Male coupler crimping with wire is damaged | Source detachment from control cable/cannot be retrieve back in the device | Potential exposure to occupational worker/public | Excessive force at crimping/wear and tear | inspection | 1 | 9 | 3 | 27 | Periodic QA testing | 14 |
| RC4 | Rotating handle | Damaged | Source cannot be projected. Source may be retracted with difficulty | Potential exposure to occupational worker (if source is in exposed position) | Accidental impact/fall | Physical verification | 2 | 5 | 3 | 30 | Periodic QA testing | 13 |
| RC5 | Projection sheath/ conduit | Damaged (from inside) | Excessive resistance is required for source assembly movement | Potential exposure to occupational worker (if source is in exposed position) | Fall of heavy object/crushing/kinking/wear & tear | No method | 3 | 5 | 9 | 135 | Method need to be developed for inspection | 05 |
| **Guide Tube** | | | | | | | | | | | | |
| GT1 | Projection sheath | Damaged (from inside) | Source stuck inside guide tube | Potential exposure to occupational worker | Fall of heavy object/crushing/kinking/wear and tear | No method | 3 | 8 | 9 | 216 | Method need to be developed for inspection | 01 |
| GT2 | | Damaged (from outside) | No effect on the operation | NAE | Fall of heavy object/crushing/ ageing | Physical verification | 4 | 3 | 2 | 24 | Periodic QA testing | 15 |
| GT3 | | Flexibility lost | Source stuck inside guide tube | Potential exposure to occupational worker | Prolonged exposure to harsh environmental conditions/ageing | Inspection | 4 | 6 | 5 | 120 | Periodic QA testing | 06 |
| GT4 | End tip of guide tube (exposure head) | Damaged/decoupled from sheath | Source may move out of the projection sheath | Potential exposure to occupational worker/public | Fall of heavy object/crushing/wear and tear | Inspection | 2 | 9 | 2 | 36 | Periodic QA testing | 11 |
| **Source Assembly** | | | | | | | | | | | | |
| SA1 | Female coupler | Crimping with wire is damaged | Female coupler part disconnected with wire. Source may be detached | Potential exposure to occupational worker | Poor crimping/wear & tear | Inspection | 2 | 8 | 9 | 144 | Stringent QC testing by manufacturer | 04 |
| SA2 | Source capsule | Damaged | Source pellets dispersion | Potential exposure to occupational worker & public (if failure occurs during source exposed condition) | Compromised material quality/ Wear & tear | Leak test (not available with user) | 1 | 9 | 10 | 90 | Material control and stringent QC testing by manufacturer | 07 |
| SA3 | | Crimping with wire is damaged | Source capsule may disconnected with wire. Source detached from assembly | Potential exposure to occupational worker/public | Poor crimping /wear and tear | Cannot be detected with active source | 2 | 9 | 10 | 180 | Stringent QC testing by manufacturer | 03 |
| **Source Housing** | | | | | | | | | | | | |
| SH1 | Pop up button/switch/ safety latch | Broken | Device Inoperable | NAE | Improper handling/ fall from height/ impact with other heavy object | Inspection/during operation | 3 | 3 | 1 | 9 | Training to the operator | 20 |
| SH2 | | Blocked/jammed | Device inoperable | NAE | No periodic maintenance | Inspection/during operation | 6 | 3 | 1 | 18 | Periodic servicing & maintenance | 17 |
| SH3 | Selector ring | Blocked/jammed | Source cannot be driven out | NAE | No periodic maintenance | Inspection/during operation | 3 | 3 | 7 | 63 | Periodic servicing & maintenance | 09 |
| SH4 | | Not rotating after control cable connection | Source cannot be driven out/device cannot be locked | NAE | Mishandling of the device/ impact with other heavy object | Inspection/during operation | 2 | 3 | 1 | 6 | Training for operation/ carefully handling of the device | 23 |
| SH5 | Source position indicator (colour indicator) | Broken/jammed/ damaged | Source location cannot be determined | Potential exposure to operator | Improper handling/ fall from height/ impact with other heavy object | Inspection/during operation | 1 | 6 | 1 | 6 | Training for operation/ carefully handling of the device | 22 |
| SH6 | | Colour(s) not visible | Source location cannot be determined | Potential exposure to operator | Wear & tear | Physical verification | 3 | 6 | 1 | 18 | Periodic servicing & maintenance | 16 |
| SH7 | Device lock | Broken/jammed/ damaged | Device inoperable | NAE | Accidental fall of the device/impact with heavy object | Inspection/during operation | 2 | 3 | 3 | 18 | Carefully handling during operation/ transport | 17 |
| SH8 | Shipping plug | Threads worn out | Device cannot be plugged from front end | Undesired exposure to operator from streaming radiations | Wear & tear | Inspection | 3 | 4 | 1 | 12 | Periodic QA testing | 18 |
| SH9 | | Cable damaged | Source positioning inside device may be marginally deviated | NAE | Wear & tear | Inspection/during operation | 1 | 1 | 7 | 7 | Periodic QA testing | 21 |
| SH10 | Safety Plug(storage cover) | Missing/Threads worn out/ damaged | Device cannot be plugged from rear end | NAE | Wear & tear/impact with other object | Inspection/during operation | 2 | 2 | 1 | 4 | Periodic QA testing | 24 |
| SH11 | Shielding structure | Damaged (visible) | Streaming of radiation | Potential exposure to occupational worker/public | Accidental fall/large impact with heavy object/crushing of the device | Visual inspection/ radiation survey of the device | 1 | 9 | 1 | 9 | Carefully handling during operation/transport | 19 |
| SH12 | | Damaged (invisible) | Streaming of radiation | Potential exposure to occupational worker/public | Accidental fall/large impact with heavy object/crushing of the device | Radiation Survey of the device | 1 | 9 | 7 | 63 | Carefully handling during operation/ transport | 08 |
| SH13 | S-tube/source tube | Damaged | Unsmooth source movement in the device/Source stuck inside device | NAE | Wear and tear | No method | 1 | 4 | 9 | 36 | Method need to be developed for inspection | 12 |

NAE= No Adverse Effect

## V. DISCUSSION AND CONCLUSIONS

Radioisotopes like Co-60, Ir-192 and Se-75 of activity range from 370 GBq to 5 TBq are used for industrial radiography applications. These isotopes are housed inside industrial radiography exposure device which provides shielding for ionizing radiation from the source. For radiography operations, these devices are manually operated by remote control unit. An accident due to equipment malfunctioning of these exposure devices may result in deterministic biological effects to the affected occupational worker and public. Design of these devices has improved with time, to improve the operational safety and to reduce the probability of accidents.

Risk assessment for the design of industrial radiography devices was carried out using FMEA methodology. The feasibility of FMEA methodology for risk assessment in the industrial radiography has been established by this study. Results are helpful to learn about necessary improvements required in current design of the exposure device for risk management. Results shows that the RPN for most of the component failure modes are well within acceptable limits considering the acceptance criteria set for the study. None of the RPN is found to be above 500, where urgent corrective actions are required. RPNs of 19 component failure modes are found to be less than 100 which falls in acceptable category. RPN of remaining 6 component failure modes were assessed in the range of 100 to 500, where corrective actions are recommended.

Out of above 6 failure modes, most serious failure is the damage of projection sheath from inside, which may result in source stuck and hence excessive exposure to occupational worker. High RPN for this failure is attributed due to high severity and lower detection probability. Fifth severe most failure is also of same nature which is due to damage of projection sheath of remote control unit. Technique to examine the inner condition of projection sheath is not available to user as well as servicing and maintenance agency. Further, it has been observed that these projection sheaths are generally continued in use, beyond their useful life, until some difficulty arises in the smooth operation of the devices. It is highly recommended to develop the technique(s) for periodic examination for the inner condition of the projection sheaths. This technique should be preferably available with user. Additionally, regulators may enforce a practice for coding (e.g. colour code, engraved marking etc.) of each projection sheath to ensure that these sheaths are not used beyond their useful life, when the probability of failure increases manifold. These suggested actions may reduce the RPN of severity rankings 1,5 &6 of table IV.

Control cable damage of remote control unit is another identified serious failure mode of radiography device. Detection of this failure mode is possible by physical inspection, but limited to the partial length of wire, which can be projection outside the sheath. Inspection of full wire length is possible by the servicing and maintenance agency. Therefore periodic inspection of control cable by the servicing and maintenance agency will be helpful in reducing its RPN.Appropriate inspection frequency may be set for this inspection. RPN severity rank 3 & 4 are associated with the damage of the crimping part of the source capsule and female coupler of source assembly respectively. Detection probability for these failure modes is very low, since it is not possible to inspect the crimping part with the active source in source assembly. It is recommended to frame and implement the policy to test the crimping part, using appropriate testing procedures of each inactive source assembly, before actual source loading. This can be adopted as part of quality control procedure at the manufacturer site or by the agency involved in source loading in the source assembly.

Most of the other failure modes can be addressed by adopting the stringent and mandatory periodic QA test procedures by the operating institutions and training to the operators. Recommended actions may be implemented in some of the selected industrial radiography institutions and this study may be repeated after specific period to analyze the effectiveness of recommendations to reduce the RPN.

Traditional FMEA technique has several reported drawbacks. However, FMEA is simple and economical method of assessment which represent useful tool to identify the areas for improvement in the design. Results of this study provides a broad picture of risk assessment for the design of industrial radiography devices. Further risk assessment studies utilizing alternative methods are recommended for industrial radiography devices and results may be compared. Identification of initiating events for accident progression have been always crucial task for risk assessment studies. Important and severe failure modes identified in this study can be utilized as initiating events for scenario development, to carry out further risk assessment studies using fault tree and event tree analysis.

## ACKNOWLEDGEMENT

Authors are thankful to all FMEA team members for providing their valuable contribution in the study. Authors are also thankful to M/s Electronic & Engineering Co. (I) P. Ltd., Mumbai, India for providing the infrastructure and technical support to conduct the study.